**Preliminary Report: Cerebral blood flow mediates the relationship between progesterone and perceived stress symptoms among female club athletes after mild traumatic brain injury.**

Yufen Chen[1], Amy A. Herrold[2], Virginia Gallagher[3], Brian Vesci[4], Jeffrey Mjannes[4], Leanne R. McCloskey[5], James L. Reilly[3], Hans C. Breiter[3]

[1]Center for Translational Imaging, Department of Radiology, Northwestern University Feinberg School of Medicine, Chicago, USA; [2]Edward Hines, Jr. VA Hospital, Hines, USA; [3]Department of Psychiatry and Behavioral Sciences, Northwestern University Feinberg School of Medicine, Chicago, USA; [4]Department of Student Health Sports Medicine, Northwestern University, Evanston, USA; [5]Department of Obstetrics and Gynecology, Northwestern University Feinberg School of Medicine, Chicago, USA

**Corresponding author:**
Yufen Chen
710 N Fairbanks Ct. LC 0-300,
Center for Translational Imaging, Neuroimaging,
Northwestern University Feinberg School of Medicine,
Chicago, IL 60611
(312) 503-2948
yfchen@northwestern.edu

**Abstract (185 words)**

Female athletes are severely understudied in the field of concussion research, despite higher prevalence for injuries and tendency to have longer recovery time. Hormonal fluctuations due to normal menstrual cycle (MC) or hormonal contraceptive (HC) use have been shown to impact both post-injury symptoms and neuroimaging measures, but have not been accounted for in concussion studies. In this preliminary study, we compared arterial spin labeling measured cerebral blood flow (CBF) between concussed female club athletes 3-10 days post injury (mTBI) and demographic, HC/MC matched controls (CON). We test whether CBF mediates the relationship between progesterone levels in blood and post-injury symptoms, which may be evidence for progesterone's role in neuroprotection. We found a significant three-way relationship between progesterone, CBF and perceived stress score (PSS) in the left middle temporal gyrus. Higher progesterone was associated with lower (more normative) PSS, as well as higher (more normative) CBF. CBF mediates 100% of the relationship between progesterone and PSS (Sobel's p-value=0.017). These findings suggest progesterone may have a neuroprotective role after concussion and highlight the importance of controlling for the effects of sex hormones in future concussion studies.

**Introduction**

Concussion is a form of mild traumatic brain injury (mTBI) that accounts for the majority of the 300,000 sports-related brain injuries among high school and college athletes in the United States each year[1]. Prior concussions can lead to increased risk to subsequent concussions [2-4] and repetitive head impacts, whether resulting in a clinically diagnosable concussion or not, can lead to cumulative functional deficits[5-7].

Neuroimaging has consistently detected changes in the brain of concussed subjects both in the acute phase as well as during recovery [8-11]. In fact, brain changes persist after post-concussion symptoms have resolved [12], suggesting neuroimaging may offer increased sensitivity to study injury recovery. Cerebral blood flow (CBF) is emerging to be a useful metric for studying concussive injury and recovery, mainly due to its tight coupling to neuronal activity and involvement in inflammatory processes [13-17]. In the acute phase of concussion, CBF is often reduced, and remains low during the recovery process [8, 12, 18, 19]. It is also significantly correlated with symptom severity and cognitive performance [10, 20], providing a more objective and quantitative complement to clinical assessment.

Collegiate athletes can be generally divided into varsity and club teams. According to the National Collegiate Athletic Association, the estimated number of varsity athletes in the US is 460,000 [21], whereas more than 2 million athletes participate in collegiate clubs [22]. Despite the much larger number of club athletes, they tend to be less studied as they receive less resources and medical supervision, as well as less education about sports-related concussion [23], even though their practice schedule and probability of exposure to head trauma are comparable to varsity athletes.

Another subgroup of athletes that are understudied are females. As of 2004-2005, approximately 41% of collegiate females participated in sports [24], which translates into roughly 200,000 female athletes. However, most concussion studies to date have focused on male athletes. Those studies that have included females have either grouped them together with males or have not accounted for sex-specific factors such as hormonal fluctuations, which may lead to inconsistent findings. There is therefore a dire need for female athlete studies, especially since female athletes appear to have a higher prevalence for injuries and tend to have longer recovery time [25-27]. A likely physiological factor accounting for these gender differences is hormones. Females experience significant fluctuations in the levels of estrogen and progesterone during the menstrual cycle (MC). These gonadal hormones have widespread non-reproductive functions in the central nervous system (CNS) [28] and have been shown to have neuroprotective effects in various disorders including ischemia, traumatic brain injury (TBI) and spinal cord injury in both animal models and humans [29-32].

Further complicating the picture is the widespread use of hormonal contraceptives (HC), which use synthetic forms of estrogen and/or progesterone to inhibit the natural cycling of their levels to prevent pregnancy. It is estimated that 4 out of 5 sexually active women use hormonal oral contraceptives, and use of other hormonal contraceptive methods have been on the rise [33]. HC users have been found to have lower post-concussive symptom severity compared to non-HC users[34, 35]. While HC use doesn't affect length of recovery, it does appear to be associated with better cognitive performance after injury[36].

Hormonal fluctuations have also been reported to influence neuroimaging outcomes, which complicates data interpretation. For example, patterns of brain activations differed across the MC for emotional processing, verbal memory and visual-spatial tasks [37-43]. Structural and connectivity differences across the MC have also been reported [44, 45]. Similarly, neurometabolite concentrations have also been reported to fluctuate across the MC and may contribute to changes after exposure to HAEs [46]. These findings highlight the importance of controlling for MC phase when studying concussion in female athletes, a methodology that has not yet been applied to this field of research.

For the current study, we investigate whether levels of progesterone, a gonadal hormone that has demonstrated neuroprotective effects in brain injury, influences post-concussive symptom severity during the acute phase (i.e. 3-10 days post-injury) in female club athletes. We incorporated a rigorous MC tracking and HC user/non-HC user matching technique with arterial spin labeling (ASL), a noninvasive neuroimaging technique for measuring CBF, to understand whether CBF mediates this relationship, which can be considered evidence for progesterone having a neuroprotective role in concussion. We used a stepwise approach to reveal brain regions that have a three-way relationship between progesterone, symptom score and CBF. Mediation analysis was then applied to determine whether there is a directed pathway for the interactions.

**Methods**

**Subjects**

At the beginning of each sports season, female athletes were approached and asked to fill out an online screener, which collects demographic information and self-report of MC start dates for the past 3 months. Injured athletes were identified by the university's Athletics department and referred to the study team for enrollment into the study after obtaining written informed consent in compliance with guidelines of the university's Internal Review Board. Study visits were divided into two consecutive days, where study assessments including: 1) Perceived Stress

Scale (PSS) [47], Post-Concussive Symptom Scale (PCSS) [48], Beck Depression Inventory II (BDI-II) [49], caffeine consumption and alcohol and cannabis use for the past 30 days were collected on Day 1, and blood sampling and MRI were completed on Day 2. The study visits were scheduled within 3-10 days post-injury. Sixteen injured athletes were identified and recruited for this study, only 15 of them completed the MRI data collection.

Sixteen control subjects were matched based on age, ethnicity, handedness and contraceptive use/type, recruited from non-collision sports teams. For MC matching purposes, all subjects were divided into 3 groups: (1) non-hormonal contraceptive users with regular MCs, (2) non-hormonal contraceptive users with irregular MCs and (3) hormonal contraceptive users. For non-hormonal contraceptive users, days 1-7 of MC were estimated as the follicular phase, and days 20 or higher were estimated as the luteal phase. Based on the 3 month timeline follow back MC tracking self-report, we scheduled the matched control subjects during the same MC phase the mTBI athlete was scanned in. For example, if the mTBI athlete was scanned during the follicular phase, the matched control was scheduled during days 1-7 of her MC. Users of hormonal contraceptives such as oral contraceptives or NuvaRing do not have normal MC as their hormones are suppressed, therefore their cycles were divided into *active* and *inactive* phases, depending on whether they were on the active hormone pills or with NuvaRing inserted, or placebo pills or NuvaRing not inserted. Control subjects were scheduled within 2 days of the matched mTBI athlete's pill pack or NuvaRing day.

**MRI Acquisition**

Imaging data were acquired on a 3.0T whole body Siemens Prisma scanner (Erlangen, Germany), using a 64-channel head/neck receive-only coil. High resolution, T1-weighted anatomical images were collected using 3D-MPRAGE with the following parameters:

TR/TE/TI/FA=2300ms/2.94ms/900ms/9°, 176 sagittal slices, 1mm isotropic resolution, iPAT acceleration factor=2, iPAT reference lines=38 (total duration=5min 37s). ASL data were collected using a 2D-EPI acquisition, using pseudo-continuous labeling [50]. Other parameters include: TR/TE=4500ms/12ms, label duration=1.5s, post-labeling delay=1.8s, labeling plane offset=90mm from center of imaging slices, resolution 3.4x3.4x6mm³, 24 slices with 1.5mm gap acquired in ascending order, iPAT acceleration factor=2, 35 pairs of interleaved control and tag images (total duration=5min 29s).

**Cerebral Blood Flow mapping**

Imaging data were processed using in-house scripts written in Matlab R2016a (Mathworks, Natick, MA) with Statistical Parametric Mapping SPM8 (Wellcome Department of Imaging Neuroscience, London, UK). All ASL data were motion-corrected with the first image of the series as the reference and then co-registered to the high resolution anatomical image. Perfusion weighted images were generated by pairwise subtraction between control and tag images and averaged over the entire time-series. Images were converted to quantitative CBF units in ml/100g/min using the single-blood-compartment model [51]:

$$f = \frac{\lambda \cdot \Delta M \cdot e^{PLD/T_{1b}}}{2\alpha \cdot M_0 \cdot T_{1b} \cdot (1 - e^{-\frac{\tau}{T_{1b}}})}$$

Where $f$ represents CBF in quantitative units, $\Delta M$ is the perfusion weighted signal, $\lambda$ is the blood/water tissue partition coefficient (assumed to be 0.9 g/ml [52]), $\alpha$ is the inversion efficiency assumed to be 0.85 [50], $M_0$ is the equilibrium magnetization from the control acquisition, and $T_{1b}$ is blood T1 assumed to be 1664ms [53]. The quantitative CBF maps were then transformed to Montreal Neurological Institute (MNI) template space and up-sampled to 1.5mm isotropic resolution based on the transformation matrix calculated from the high-resolution anatomical image using VBM8 [54].

Given the poor spatial resolution of the ASL acquisition, partial volume correction (PVC) is necessary to minimize contamination of ASL signal from different tissue types. Tissue probabilities from segmented gray and white matter maps of the high-resolution anatomical image were used to calculate true gray matter CBF based on the following equation, where the GM flow was assumed to be 2.5 times that of WM flow [55]:

$$f_{corr} = \frac{f_{uncorr}}{(P_{GM} + 0.4 * P_{WM})}$$

To minimize artifacts due to low GM probability, only voxels with at least 30% GM were corrected. Voxels with less than 30% GM were set to 0. PVC maps were smoothed with 8mm FWHM kernel before entering into statistical analysis.

**Blood Draw**

For assessment of progesterone level, 1mL of blood was collected by a trained nurse of the Northwestern Memorial Hospital, outpatient Clinical Research Unit, and analyzed at the Northwestern Memorial Hospital Clinical lab.

**Data Analysis**

Given the pilot sample size and multiple symptom scores, we sought to increase our statistical power by combine the PCSS, PSS and BDI scores into a composite score (CompositeZ). For each of these scores, a z-score was computed by subtracting the group mean of the controls from each subject's score and dividing by the standard deviation of the controls. The composite score was calculated as the average of the 3 z-scores. Standard 2-sampled t-tests were used to determine group differences in the symptom and composite scores.

As the main focus of this study was to determine if there were any 3-way relationships between regional CBF, progesterone levels and symptom scores, we first computed Pearson's correlation coefficients between symptom scores and progesterone levels to decide which symptom score to use for the imaging-based statistics.

A stepwise process was used for the imaging-based analysis in Statistical Parametric Mapping (SPM8, Wellcome Institute). We first identified regions in the brain where the mTBI and CON groups differed in CBF. This was performed with a voxelwise two-sampled t-test, with each subject's global GM CBF used to proportionally scale each subject's CBF map to a global value of 50 mL/100g/min. Proportional scaling is necessary to eliminate variability arising from each subject's global CBF level. Given the exploratory nature of this analysis on a small sample, we used a liberal threshold of $p=0.05$, cluster of 10 voxels.

Within areas with significant group differences in CBF, we then used voxelwise multiple regression to identify regions where there was a significant correlation between symptom score and CBF. Each subject's global CBF level and the number of days from injury to imaging were included as covariates. Contrasts for positive and negative correlations with symptom score were thresholded at $p<0.005$, cluster of at least 20 voxels. No correction for multiple comparison was used due to the exploratory, stepwise nature of this analysis. Regional CBF (rCBF) values were extracted from significant clusters, scaled by the scale factor calculated from proportional scaling, and used to compute Pearson's correlation coefficients by correlating rCBF to progesterone levels.

Clusters where rCBF significantly correlated with both symptom score and progesterone levels were entered into the mediation analysis, which included 1) a directed analysis, where progesterone was designated as the independent variable (IV) and cluster rCBF the mediator (M), and 2) a control analysis where cluster rCBF was designated as the IV and progesterone the M. In both analyses, the symptom score was the dependent variable. The following criteria need to be satisfied for a mediation relationship to be considered significant: 1) significant 3-way relationships between IV, M, and DV (i.e. paths A, B and C must all be statistically significant with p<0.05), 2) when mediator variable was added to regression between IV and DV, the standard beta coefficient was reduced and the regression was no longer significant, 3) direct mediation Sobel's p-value < 0.05 and control mediation Sobel's p-value > 0.05, and 4) % effect mediated for direction mediation should be greater than that for the control mediation. Mediations analyses were conducted in SPSS Version 26 and utilizing the mediation testing from MacKinnon and Dwyer [56]. The first criteria mentioned above is essentially a conjunction analysis with a p-value threshold $p_a*p_b*p_c$, where $p_{subscript}$ represents the p-value for that pathway. By requiring each of the pathways to have a p-value of 0.05 or less, the conjunction threshold p-value is therefore 0.05*0.05*0.05=0.000125. For each mediation analysis, we compared the conjunction p-value to the Bonferroni corrected p-value of .00136 (.05/44), which is based on [ (3 areas where CBF was significantly related to progesterone, Path A) + (4 areas where CBF was significantly related to PSS_ZCon, Path B) + (4 symptom score relationships to progesterone levels tested, Path C)] x 2 for mTBI and CON x 2 for directed mediation hypothesis testing where progesterone is the IV and CBF is the mediator and for control mediation testing where CBF is the IV and progesterone is the mediator.

**Results**

A summary of the subject demographics, symptom scores and progesterone levels is shown in Table 1. Of the original sample of 15 mTBI and 15 control athletes that completed the MRI, one mTBI athlete did not have usable ASL data and was excluded with her matched control. Another pair of mTBI/control athletes was assessed during the luteal phase of their MC, where their progesterone levels were over 10 times higher than the rest of our cohort. Given the small sample size of this study, this pair of athletes also had to be excluded from the analysis, resulting in a final sample of 26 club female athletes (mTBI n=13, control n=13). Among the symptom scores including the composite score, all except the PCSS emotional subscore and BDI were significantly higher in the mTBI group.

**Table 1.** Summary of demographics, symptoms and outcomes.

|  | mTBI (n=13) | CON (n=13) | p-value |
|---|---|---|---|
| Age (M ± SD) | 20.2 ± 1.5 | 20.1 ± 1.1 | n.s. |
| N Hormonal contraceptive users | 7 | 7 | n.s. |
| PCSS | 21 ± 15 | 4 ± 4 | 0.003 |
| PCSS sleep | 4 ± 4 | 1 ± 1 | 0.01 |
| PCSS emotional | 2 ± 2 | 1 ± 2 | n.s. |
| PCSS cognitive | 5 ± 4 | 0 ± 1 | 0.002 |
| PCSS physical | 10 ± 7 | 2± 2 | 0.001 |
| PSS | 15 ± 6 | 8 ± 5 | 0.002 |
| BDI | 4 ± 6 | 1 ± 1 | n.s. |
| CompositeZ | 2.09 ± 2.59 | -0.22 ± 0.64 | 0.008 |
| Return to Play (days) | 22 ± 14 | - | - |
| Days to symptom baseline | 1 ± 5 | - | - |
| Days from injury to scan | 7 ± 2 | - | - |
| Progesterone (ng/mL) | 0.48 ± 0.27 | 0.43 ± 0.32 | n.s. |

M ± SD, Mean ± standard deviations; n.s., group comparisons are not significant at p<0.05 level.

Table 2 shows the Pearson's correlation coefficients (p-values) for groupwise correlations between individual symptom Z scores and the composite Z score with progesterone levels. Only PSS_ZCon was significantly correlated with progesterone levels in the mTBI group (r = -0.586, p = 0.035), such that as progesterone levels are higher, perceived stress symptoms are lower or less severe. Based on this finding, all subsequent imaging and mediation results are reported in reference to PSS_ZCon.

**Table 2**. Pearson's correlation coefficients (p-values) between symptom scores and progesterone levels.

| Symptom Score | mTBI | CON |
|---|---|---|
| PCSS_ZCon | -0.294 (0.329) | -0.007 (0.983) |
| PSS_ZCon | **-0.586 (0.035)** | -0.3 (0.319) |
| BDI_ZCon | -0.327 (0.275) | 0.09 (0.769) |
| Composite_ZCon | -0.397 (0.179) | -0.1 (0.745) |

Symptom scores were converted into Z scores based on the mean symptom scores of the CON group. Correlation between PSS_ZCon of mTBI group and progesterone level was the only one significant at p<0.05, highlighted in bold.

Figure 1 shows clusters with significant correlation between regional CBF and PSS_ZCon, overlaid onto a single subject's high resolution anatomical image in MNI template space. All images are in neurological convention, where left side of the brain is on the left side of the image. Red clusters represent positive correlations, i.e. higher symptoms associated with higher rCBF, and blue clusters represent negative correlations. A summary of the clusters, including number of voxels for each cluster, MNI coordinates of the voxel with the highest T-value, and anatomical labels based on the Harvard-Oxford cortical (48 regions) and subcortical (21 regions) atlases [57] and a probabilistic cerebellar atlas with 28 anatomical regions [58] are shown in Table 3. Four clusters had significant correlations between rCBF and progesterone for the mTBI group only: left superior parietal lobule (L. SPL), left medial temporal gyrus (L. MTG), superior left superior temporal gyrus (L. STG), and left precuneus. While the L. SPL cluster had a positive correlation between rCBF and PSS_ZCon (i.e. higher rCBF associated with higher or more severe PSS_ZCon), the remaining 3 clusters had negative correlations between rCBF and PSS_ZCon (lower rCBF associated with higher or more severe PSS_ZCon). No significant correlations to progesterone were found for the CON group in any of the clusters listed in Table 3. Three of these clusters had significant correlations between rCBF and progesterone levels, again only in the mTBI group. The correlation coefficients and p-values between rCBF and progesterone levels for these clusters are shown in bold in Table 3.

**Figure 1**. Clusters where CBF in mTBI group correlated with PSS_ZCon score, overlaid onto a single subject's T1 in MNI space. Red represents positive correlation, i.e. higher PSS_ZCon score associated with higher CBF, and blue represents negative correlation, i.e. higher PSS_ZCon score associated with lower CBF. Clusters were generated by thresholding the results at p<0.005, at least 20 voxels for each contrast (positive and negative correlations with PSS_ZCon score), and only in regions where a significant group difference in CBF was detected.

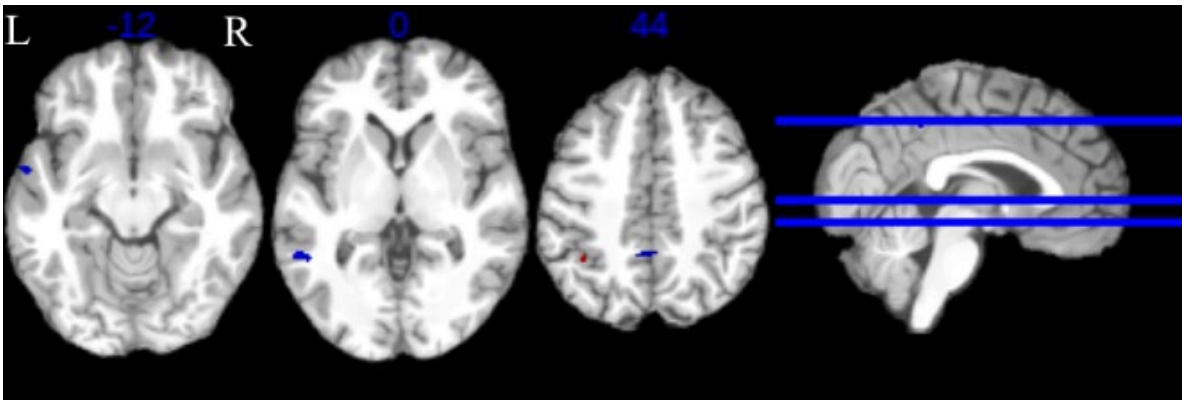

Table 3. Summary of clusters with significant correlation between CBF and the PSS_ZCon score in the mTBI group, masked by areas with significantly different CBF between mTBI and CON groups. Clusters were generated using a p-value threshold of 0.005 and cluster threshold of 20 voxels. The two rightmost columns show Pearson's correlation coefficients and p-values in parentheses for correlations between CBF and progesterone levels for each group. Correlations with p-value < 0.05 are shown in **bold.**

| *+PSS, days to scan, proportional* | | | | | | | |
|---|---|---|---|---|---|---|---|
| Nvoxels | T | equivZ | p(unc) | x,y,z (mm) | anatomical label | mTBI vs. Prog | CON vs. Prog |
| 21 | 5.534 | 3.663 | 0.0001 | -36, -51, 46 | Left SPL (59%), post. Left Supramarginal (26%) | **-0.562 (0.046)** | 0.028 (0.929) |
| *-PSS, days to scan, proportional* | | | | | | | |
| Nvoxels | T | equivZ | p(unc) | x,y,z {mm} | anatomical label | mTBI vs. Prog | CON vs. Prog |
| 42 | 4.812 | 3.385 | 0.0004 | -54, -48, 1 | Left MTG, temporoocc.(80%), OUTSIDE (18%), post. Left MTG (2%) | **0.713 (0.006)** | 0.023 (0.94) |
| 27 | 4.179 | 3.107 | 0.0009 | -58, -3, -11 | Sup. Left STG (67%), ant. Left MTG (26%) | **0.654 (0.015)** | 0.228 (0.455) |
| 32 | 3.767 | 2.905 | 0.0018 | -8, -48, 42 | Left precuneus (59%), Right precuneus (27%), post. Left Cingulate (12%), post. Right Cingulate (2%) | 0.506 (0.077) | -0.286 (0.343) |

Abbreviations: Nvoxels, number of voxels in cluster; cereb., cerebellum; post., posterior; temporoocc, temporooccipital; ITG, Inferior Temporal Gyrus; MTG, Middle Temporal Gyrus; MFG, Middle Frontal Gyrus; SFG, Superior Frontal Gyrus; STG, Superior temporal gyrus; SPL, Superior Parietal Lobule; ant., anterior; sup., superior; SMA, Supplementary Motor Cortex. OUTSIDE means the cluster was not in a labeled region of the atlas.

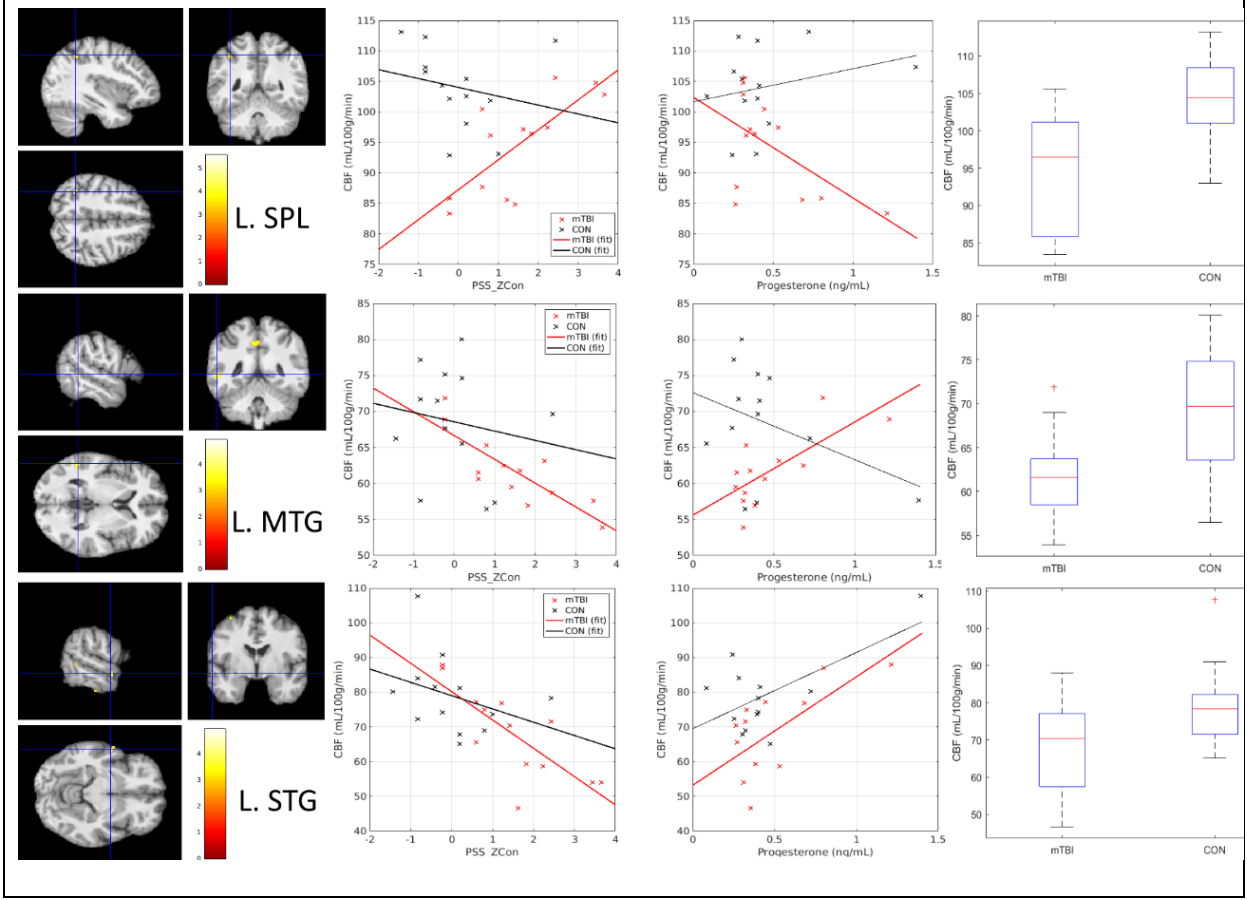

Figure 2. Clusters where there were significant correlations between CBF and progesterone levels for the mTBI group. The corresponding CBF-PSS_ZCon and CBF-progesterone plots are shown to the right, where red represents mTBI group and black represents CON group. The boxplots in the rightmost column show the distribution of CBF values extracted from this cluster for the two groups. The red line represents the group mean CBF, and outliers are denoted by a red cross.

The locations of the 3 clusters with significant three-way relationships between rCBF, PSS_ZCon and progesterone, and the corresponding two-way scatterplots are shown in Figure 2, with mTBI group in red and CON group in black. In the L. SPL cluster which had a positive correlation between rCBF and PSS_ZCon, the correlation between rCBF and progesterone was negative. Similarly, in the clusters with a negative correlation between rCBF and PSS_ZCon, the correlation between rCBF and progesterone was positive. None of the correlations was significant for the CON group. In all 3 clusters, mean CBF for the mTBI group was lower than that of the CON group ($p<0.05$).

Mediation results are shown in Table 4, where odd-numbered rows represent the directed mediations, with progesterone as IV and rCBF as M, and even-numbered rows represent the control mediations, with rCBF as IV and progesterone as M. In all the directed mediation cases, paths A, B and C without M (model 1) were statistically significant, but after including the effects of M, path C (model 2) was no longer significant, suggesting rCBF is an effective mediator of the relationship between progesterone and PSS_ZCon. In all 3 clusters, the % effect of progesterone on PSS_ZCon mediated by rCBF were greater than the % effect of rCBF on PSS_ZCon mediated by progesterone, suggesting that only the directed mediation is valid. The L. MTG cluster, denoted by *, was the only cluster with a Sobel's test p-value less than 0.05 and mediates 100% of the neuroprotective effect of progesterone on PSS_ZCon. Correcting for multiple comparisons, each conjunction analyses p-value of paths A, B, and C ($p_a*p_b*p_c$) was less than the Bonferroni correction $p = .05/44 = .00136$ (rightmost column of Table 4). The location of this cluster, as well as scatterplots for paths A, B and C model 1 are shown in Figure 3.

Table 4. Mediation results for mTBI group, using CBF values extracted from the 3 ROIs where a significant correlation between CBF and progesterone were detected in Table 3. In all 3 ROIs, CBF mediates a higher percentage of the effect of the IV (progesterone) on DV (PSS_ZCON), compared to progesterone as a mediator for the relationship between CBF and PSS_ZCON. Sobel's test p-value of < 0.1 was considered statistically significant and denoted with *. Correcting for multiple comparisons, each conjunction analyses p-value ($p_a*p_b*p_c$) was less than the Bonferroni correction p (Bonf) = .05/44 = .00136.

| Path: Model & Predictor(s): | | | Path A: IV predicting M | | Path B: M predicting DV | | Path C model 1: IV predicting DV no M | | Path C model 2: IV predicting DV with M | | % Effect Mediated | Sobel Test p-value | p(Bonf) = .00136 |
|---|---|---|---|---|---|---|---|---|---|---|---|---|---|
| IV | M | DV | Std β | p-value | Std β | p-value | Std β | p-value | Std β | p-value | | | $p_a*p_b*p_c$ |
| progesterone | L. SPL | PSS_ZCON | -0.566 | 0.044 | 0.748 | 0.003 | -0.585 | 0.036 | -0.238 | 0.35 | 59 | 0.091 | 4.752E-6 |
| L. SPL | progesterone | PSS_ZCON | -0.566 | 0.044 | -0.585 | 0.036 | 0.748 | 0.003 | 0.613 | 0.03 | 18 | 0.368 | 4.752E-6 |
| progesterone | L. MTG | PSS_ZCON | 0.728 | 0.005 | -0.826 | 0.001 | -0.585 | 0.036 | 0.034 | 0.898 | 100 | 0.017* | 1.80E-7 |
| L. MTG | progesterone | PSS_ZCON | 0.728 | 0.005 | -0.585 | 0.036 | -0.826 | 0.001 | -0.851 | 0.008 | 0 | 0.896 | 1.80E-7 |
| progesterone | L. STG | PSS_ZCON | 0.664 | 0.013 | -0.772 | 0.002 | -0.585 | 0.036 | -0.13 | 0.636 | 78 | 0.052 | 9.36E-7 |
| L. STG | progesterone | PSS_ZCON | 0.664 | 0.013 | -0.585 | 0.036 | -0.772 | 0.002 | -0.685 | 0.028 | 11 | 0.630 | 9.36E-7 |

Figure 3. Correlation plots of the 3-way relationship between progesterone, rCBF and PSS_ZCon of the L. MTG cluster, which had the strongest mediation effect.

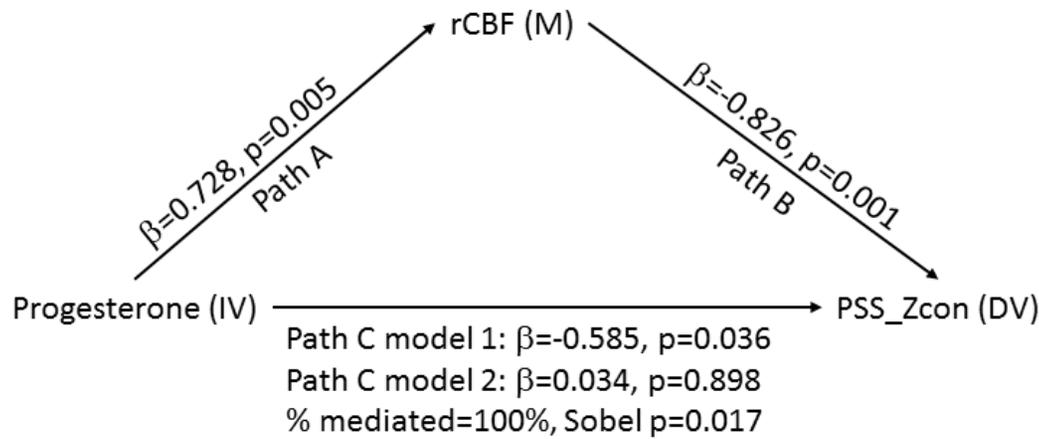
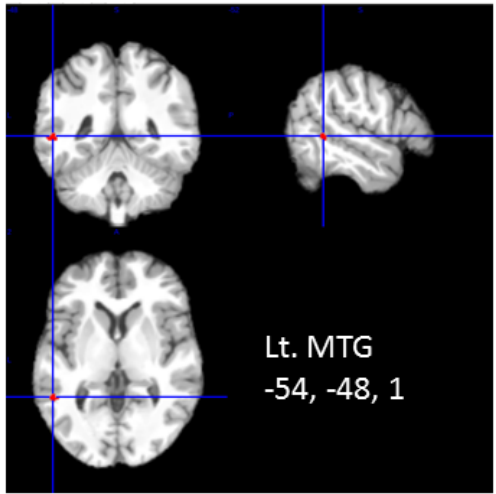
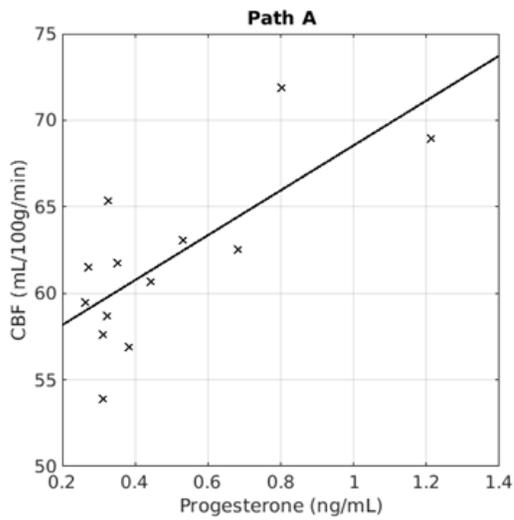
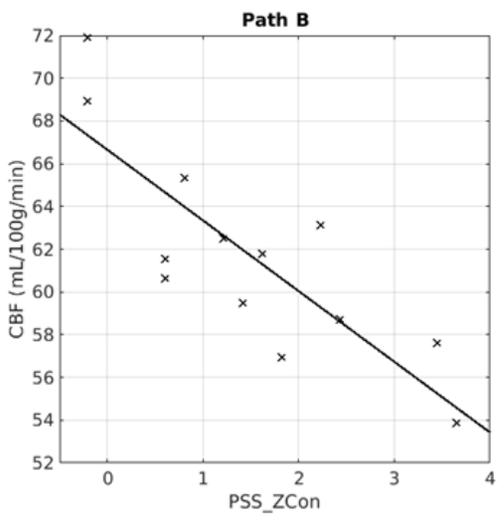
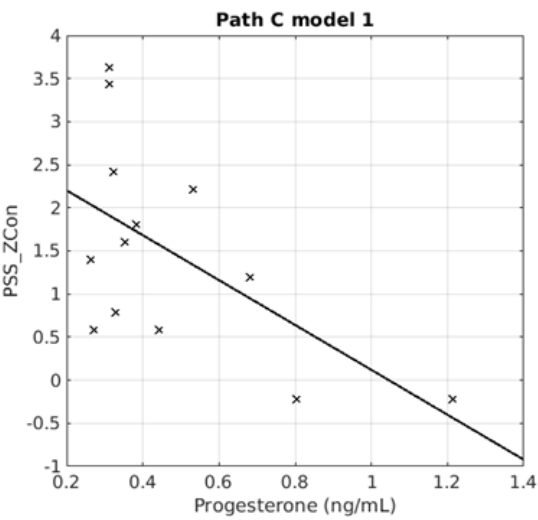

**Discussion**

In this pilot study, we investigated if progesterone, CBF and post-concussive symptoms are related in female collegiate club athletes assessed 3-10 days after sports-related concussion, compared to demographic, HC, and MC matched control athletes participating in non-contact sports. Our goal was to identify areas of the brain where CBF was significantly related to both progesterone levels and symptom outcomes in order to test this three-way relationship in a mechanistic mediation analysis, which may be construed as evidence for possible neuroprotective effects of progesterone in concussion. While various symptom scores including PCSS, a subset of its subscores, and PSS were all elevated in the mTBI group, only PSS was significantly correlated with progesterone levels in the mTBI group in a negative relationship, i.e. higher progesterone levels were associated with lower PSS scores. Using voxelwise statistical analysis, we found 3 brain regions, all localized in the left hemisphere, with a three-way relationship between progesterone, CBF and PSS: left SPL, left MTG and left STG. Mediation analysis revealed only the left MTG cluster was statistically significant, based on Sobel's test p-value of 0.017. rCBF mediated 100% of the relationship between progesterone and PSS in this cluster.

Progesterone is a gonadal hormone that is synthesized by the ovaries in females, and testes and the adrenal cortex in males [32]. Progesterone mediates non-reproductive functions in the central nervous system via an array of progesterone receptors that are widely distributed in the brain [59]. Various animal models of brain injury have demonstrated that progesterone has neuroprotective effects. In rats, progesterone administered before middle cerebral artery occlusion resulted in smaller areas of infarction and better outcome [60]. It also improved functional measures when administered after stroke [61, 62]. Progesterone also has neuroprotective effects in a rodent model of TBI, where levels of lipid peroxidation, cerebral edema and inflammatory proteins associated with brain damage were all reduced after progesterone administration [31, 63]. Clinical trials in adult TBI patients have also reported positive outcomes associated with progesterone treatment including reduced death rate and improved functional outcomes [64-66].

In our cohort, we found higher progesterone levels associated with lower PSS score, i.e. more normative levels. Concurrently, higher progesterone was also associated with higher rCBF values in the left MTG. Higher CBF has traditionally been associated with better tissue health and cognitive performance [67-69], so this finding suggests higher progesterone levels are associated with better tissue health. Our between group data demonstrating that CBF levels were lower for the mTBI relative to control athletes also suggest that higher CBF values are more normative. Further, we showed with mediation analysis that CBF mediates the relationship between progesterone and PSS, which can be viewed as evidence that progesterone has a neuroprotective role and that this in turn results in reduction of post-concussion symptom.

The locus of this mediation relationship is a small cluster within the posterior part of the left MTG. The MTG is a known network hub that facilitates communication between parallel and distributed brain networks [70, 71]. The posterior part of the MTG is an important center for semantic processing, which has been shown to be slightly left lateralized based on a meta-analysis of 120 studies on semantic processing [72]. This is a heteromodal region that is activated by both visual and auditory stimuli and is thought to integrate various types of stimuli and control semantic retrieval [72, 73]. While it is not immediately obvious how the left MTG is associated with stress, there's evidence this region is also involved in emotional face processing

[74] and implicated in social anxiety disorder, where functional connectivity of this network hub is impaired and proportional to both clinical and patient-reported symptom severity [75].

The current finding warrants further investigation as the relationship between progesterone and CBF is not widely studied. Although CBF in females fluctuates over the course of the menstrual cycle during which the balance between estrogen and progesterone varies greatly, the exact mechanism is unclear [76]. Different types of progestin in hormone replacement therapy for menopausal women also seem to play a role in modulating CBF [77]. A natural extension of this study would be to include measures of estrogen in order to get the full picture, as estrogen and progesterone have been shown to work in an antagonized manner. Estrogen also has widely studied effects on CBF, primarily through the interaction between estrogen receptors and endothelial nitric oxide synthase [78].

The most obvious limitation of the current study is the small, pilot sample size, which limited our analysis to subjects within the follicular phase of the MC. A larger sample size will provide adequate statistical power to study both progesterone and estrogen across multiple phases of the MC to improve understanding of how the balance of these hormones might have neuroprotective effects. Furthermore, a larger sample size including more HC users may also help elucidate whether HC use is also related to CBF and post-injury outcomes. Another potential weakness is related to the nonspecific nature of the PSS score. PSS assesses a person's ability to handle stress, which can vary even in healthy subjects without concussion. Armed with the current findings in the left MTG, which has a crucial role in semantic processing, future studies can use more specific tasks to determine if semantic processing is affected in concussed subjects. Finally, while our MC/HC use matching protocol helps to minimize measurement bias due to the effect of MC phases on neuroimaging measures, it does not allow us to determine whether hormone levels at time of injury also play a role in affecting symptoms and recovery. Inclusion of this information in future studies can help improve our understanding of the multi-way relationship between hormones, neuroimaging measures and post-injury symptoms.

In conclusion, we present results from a preliminary study examining how CBF factors into the neuroprotective effects of progesterone in a group of collegiate club female athletes after concussion. We found high progesterone levels associating with lower or more normative PSS scores, as well as higher rCBF in the left MTG, reflecting potential neuroprotective effects of progesterone. rCBF in the left MTG mediates 100% of progesterone's relationship with PSS, which could be interpreted as the underlying mechanism. These findings warrant further investigation in a large-scale study to better understand the role of multiple hormones, both natural and synthetic, in influencing post-concussion symptoms by their actions in the brain.